\documentclass[aps,prl,floatfix,twocolumn,a4paper,showpacs]{revtex4}
\pdfoutput=1
\usepackage{graphicx}
\usepackage{tikz}
\usepackage{amsmath}
\usepackage{amssymb}
\usepackage{epsf,latexsym}
\usepackage{times}
\usepackage{xspace}

\newcommand{\PAPER}{paper\xspace}

\newcommand{\half} {\ensuremath{\frac{1}{2}}}

\newcommand{\ket}[1]{\ensuremath{\left| #1 \right\rangle}}

\newcommand{\EX}[1] {\ensuremath{\left\langle #1 \right\rangle}}
\newcommand{\be}{\begin{equation}}
\newcommand{\bel}[1]{\begin{equation}\label{#1}}
\newcommand{\bal}[1]{\begin{eqnarray}\label{#1}}
\newcommand{\ee}{\end{equation}}
\newcommand{\ba}{\begin{eqnarray}}
\newcommand{\ea}{\end{eqnarray}}
\newcommand{\ud}{\mathrm{d}}
\newcommand{\ui}{\mathrm{i}}
\newcommand{\Fig}[1]{Fig.~\ref{#1}}
\newcommand{\Eq}[1]{Eq.~(\ref{#1})}

\newcommand{\cVector}[2]{ \begin{bmatrix}#1 \\ #2 \end{bmatrix}}
\newcommand{\rVector}[2]{ \begin{bmatrix}#1 & #2 \end{bmatrix}}
\newcommand{\Matrix}[4]{\begin{bmatrix}#1 &#2\\ #3 & #4\end{bmatrix}}

\newcommand{\MJE}{M.J. Everitt}
\newcommand{\JHS}{J.H. Samson}
\newcommand{\SSa}{S.E. Savel'ev}
\newcommand{\RW}{R. Wilson}
\newcommand{\AMZ}{A.M. Zagoskin}
\newcommand{\TPS}{T.P. Spiller}

\newcommand{\LEEDS}{Quantum Information Science, School of Physics and Astronomy, %
University of Leeds, Leeds LS2 9JT, UK.}
\newcommand{\LBRO}{Department of Physics, Loughborough University,
                Loughborough, Leics LE11 3TU, United Kingdom}

\begin{document}

\usetikzlibrary{arrows,shapes,positioning}
\usetikzlibrary{decorations.markings}

\usetikzlibrary{backgrounds}
\tikzstyle arrowstyle=[scale=1]
\tikzstyle directed=[postaction={decorate,decoration={markings,
    mark=at position .65 with {\arrow[arrowstyle]{stealth}}}}]
\title{Tunable refraction in a two dimensional quantum metamaterial}

\author{\MJE}
\email{m.j.everitt@physics.org}
\affiliation{\LBRO}
\author{\JHS}
\affiliation{\LBRO}
\author{\SSa}
\affiliation{\LBRO}
\author{\RW}
\affiliation{\LBRO}
\author{\AMZ}
\affiliation{\LBRO}
\author{\TPS}
\affiliation{\LEEDS}

\date{\today}

\begin{abstract}
In this \PAPER we consider a two-dimensional metamaterial comprising an array of qubits (two level quantum objects).  Here we show that a two-dimensional quantum metamaterial may be controlled, e.g. via the application of a magnetic flux, so as to provide controllable refraction of an input signal. Our results are consistent with a material that could be quantum birefringent (beam splitter) or not dependent on the application of this control parameter. We note that quantum metamaterials as proposed here may be fabricated from a variety of current candidate technologies from superconducting qubits to quantum dots. Thus the ideas proposed in this work would be readily testable in existing state of the art laboratories.
\end{abstract}



 
\maketitle

Quantum metamaterials, i.e., artificial optical media, which maintain quantum coherence over the signal traversal time, hold promise of becoming a testing ground for the investigation of the quantum-classical transition, interesting new phenomena in wave propagation, and unusual technological applications~\cite{Felbacq,Pile}. 
With strong analogies existing between atomic physics, quantum optics and superconducting systems it is natural to seek technologies that span these fields~\cite{You}. Solid state quantum metamaterials is one such class of devices where such parallels can be leveraged to great utility.
Indeed, emphasising such a synergy, the implementation of a quantum metamaterial in the optical range is feasible~\cite{Felbacq,Quatch}. In our view, the experimental realisation of the concept is likely to be achieved first in the microwave range, as was the case with conventional metamaterials~\cite{Pendry}. We believe that the best candidate system system would comprise superconducting qubits~\cite{Rakhmanov,Zagoskin} playing the role of controllable artificial atoms. This view is supported by the ability of superconducting flux qubits to play the role of  quantum scatterers - a phenomena that has been both theoretically modelled  and experimentally observed~\cite{Astafiev-Science,Astafiev-PRL,Abdumalikov}.

One dimensional quantum metamaterials can be realised readily enough. An example of this is a chain of qubits placed in a transmission line~\cite{Rakhmanov,Zagoskin,Astafiev-Science}. Unfortunately, 1D-devices do not allow us to realise more interesting and useful effects such as ``quantum birefringence'' and other phenomena, where the change of direction of the signal by an arbitrary angle is important (e.g., ~\cite{Quatch}). Here one needs to go beyond 1D. 
As with any truly quantum circuit, we retain the essential requirement  that any ``proof of principle'' realisation of a truly quantum metamaterial must maintain global quantum coherence, and it is therefore necessary for such a system to contain as few unit elements as possible. In this work we are therefore concerned with the minimal realisation of a 2D quantum metamaterial.  

The effect of changing signal transmission amplitude through a system by tuning the quantum states of the constituents, which we consider in this paper, is related to two classes of phenomena: mesoscopic transport in quantum interferometers (for a review see, e.g.,Ch. 5 of ~\cite{Imry2008} or Ch.3 of ~\cite{Zagoskin})  and electromagnetically induced transparency [EIT], effect known in optics~\cite{RevModPhys.77.633,Harris1997,PhysRevLett.101.253903} and recently  demonstrated in an artificial atom (superconducting qubit) in the microwave range~\cite{Abdumalikov}. The underlying physics of either of these phenomena is the same: quantum state-dependent interference between different quantum trajectories contributing to the probability amplitude of the signalÕs transmission through the system. 
In the case of EIT, in the presence of one (``coupling'') laser field  destructive interference occurs between different transitions in an atom under the influence of the probe laser field; as a result, the field effectively does not interact with the atom.  The control is exercised through the coupling laser field, which produces the appropriate state of the atom; the mechanism works for each atom in the medium independently.

In case of mesoscopic transport through a quantum interferometer, the transmission amplitude is determined by the interference of electron wave components propagating through the branches of the device. Here the interference pattern occurs in real space, and it can be directly affected by the electromagnetic field acting on the electrons Ð most spectacularly, through Aharonov-Bohm or Aharonov-Casher mechanisms~\cite{Aharonov1984}.  

The propagation of electromagnetic signals through a planar quantum metamaterial, a toy model of which we are considering here, is determined by quantum interference in real space, like in mesoscopic transport. On the other hand, one cannot directly affect the phase of the electromagnetic wave at any reasonable field magnitude; instead, like in EIT, we control the coupling between, and thus setting the ``preferred'' states of, the artificial atoms, which constitute our medium (with the advantage that here Ð in principle Ð quantum state of each of them can be controlled individually and at will).

Systems of coupled spin-$\tfrac{1}{2}$ particles have been widely studied in a range of different physical contexts. Recently, there has been much interest in the potential use of spin chains or networks as buses for quantum state transfer within quantum information processing devices \cite{Bose2007,Kay2011}. It has been shown that the phase shift caused by an applied field (the Aharonov-Casher effect \cite{Aharonov1984}) can significantly enhance the maximum attainable degree of entanglement in a spin chain, as well as improving the transfer of entanglement around a ring of spins \cite{Maruyama2007}. The effects of continuously monitoring the output nodes of a spin network was investigated in \cite{Shizume2007} and it was demonstrated that high-fidelity quantum state transfer is  possible. We note  the emphasis of~\cite{Shizume2007}  differs from ours as they were concerned with communicating a particular state while we are interested in characterising a particular material. Indeed, we do not ask questions about the actual state of the system as a whole which will mostly be some entangled state of the qubits; we are simply interested in an output signal comprising local measurements on two edges of the material.

\begin{figure}[!tb]
\begin{center}
\begin{tikzpicture}[scale=0.6]
\node[circle, shading=ball, ball color=magenta,label=-135:$L_\mathrm{1}\propto \sigma^{+}$] (a) at (0,0) {1};
\node[circle, shading=ball, ball color=cyan] (b) at (3,0) {2};
\node[circle,draw, shading=ball, ball color=green,label=right:$L_\mathrm{3}\propto \sigma^{-}$] (c) at (6,0) {3};

\node[circle, shading=ball, ball color=cyan] (d) at (0,3) {4};
\node[circle, shading=ball, ball color=cyan] (e) at (3,3) {5};
\node[circle,draw, shading=ball, ball color=green,label=right:$L_\mathrm{6}\propto \sigma^{-}$] (f) at (6,3) {6};

\node[circle,draw, shading=ball, ball color=green,label=above:$L_\mathrm{7}\propto \sigma^{-}$] (g) at (0,6) {7};
\node[circle,draw, shading=ball, ball color=green,label=above:$L_\mathrm{8}\propto \sigma^{-}$] (h) at (3,6) {8};
\node[circle,draw, shading=ball, ball color=green,label=45:$L_\mathrm{9}\propto \sigma^{-}$] (i) at (6,6) {9};

\draw[blue,directed,ultra thick] (a) -- (b) ;
\draw[double, color = black] (b) -- (c) ;

\draw[blue,directed,ultra thick] (d) -- (a) ;
\draw[blue,directed,ultra thick] (b) -- (e) ;
\draw[double, color = black] (c) -- (f) ;

\draw[blue,directed,ultra thick] (e) -- (d) ;
\draw[double, color = black] (e) -- (f) ;

\draw[double, color = black] (d) -- (g) ;
\draw[double, color = black] (e) -- (h) ;
\draw[double, color = black] (f) -- (i) ;

\draw[double, color = black] (g) -- (h) ;
\draw[double, color = black] (h) -- (i) ;
\end{tikzpicture}
\end{center}
\caption{(color online) Set up, each node represents a qubit and each edge the qubit qubit coupling coupling. The directed couplings correspond to $\Xi_{ij}(\phi)=\sigma_{i}^{+}\sigma_{j}^{-}\exp[\ui\phi]+\sigma_{i}^{-}\sigma_{j}^{+}\exp[-\ui\phi]$ and the undirected couplings to $\Xi_{ij}(0)$.}
\label{fig:schematic}
\end{figure}
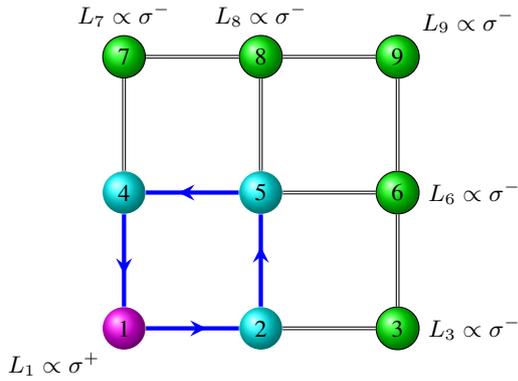
Our simulated quantum metamaterial comprises an array of interacting qubits as depicted in~\Fig{fig:schematic}. In this work we consider the reduced model for qubits coupled to electromagnetic modes where the field  degrees of freedom have been eliminated producing effective couplings between the qubits similar to~\cite{RevModPhys.73.357}. The Hamiltonian for this system is then 
\bel{eq:ham}
H=\sum_{i}\frac{1}{2}\sigma^{z}_{i}+\sum_{(i,j)\in A}\mu_{ij} \Xi_{ij}(\phi_{ij})
\ee
where $\sigma^{z}_{i}$ is the Pauli matrix for the $z$ direction and we have chosen $\mu_{ij}=\half$ for the connected qubits as illustrated through the use of edges in~\Fig{fig:schematic} and zero otherwise, i.e. $A=\{(1,2), (2,5), (5,4), (4,1), (2,3), (5,6), (5,8), (4,7), (7,8),\\ (8,9), (9,6), (6,3)\}$. The coupling operator $\Xi_{ij}(\phi_{ij})$ takes the form:
\bel{eq:coupling}
\Xi_{ij}(\phi_{ij})=\sigma_{i}^{+}\sigma_{j}^{-}\exp[\ui\phi_{ij}]+\sigma_{i}^{-}\sigma_{j}^{+}\exp[-\ui\phi_{ij}].
\ee
where $\sigma_{i}^{\pm} = \frac12 \left(\sigma_{i}^{x} \pm \ui \sigma_{i}^{y}\right)$. In order to model the effect of (for example) a magnetic field on this qubit metamaterial we have chosen the control parameter, $\phi_{ij}=\phi$ for the directed couplings $(1,2)$, $(2,5)$, $(5,4)$, $(4,1)$ and zero otherwise as indicated in~\Fig{fig:schematic}. The initial state of the system was chosen to be $\bigotimes_{i}\ket{-1}_{i}$ where $\ket{-1}_{i}$ is the eigenstate of $\sigma^{z}_{i}$ with eigenvalue $-1$. 
We note that for our candidate realisation of superconducting qubits the phase $\phi$ may be switched at high frequencies, e.g. of the order of a few GHz, ~\cite{Plantenberg}. Consequently, rapid control of a quantum metamaterial such as the one proposed here should be possible. As it may be of utility when considering classical analogues of this system we also note, that it is  possible to write the Hamiltonian in the following form:
\be \nonumber
H  =    \sum_{i} {\sigma_{i}^{z}\over 2} +  \sum_{<ij>} \frac{\mu_{ij}}{2}
\rVector{\sigma_{i}^{x}}{\sigma_{i}^{y}}
\Matrix {\cos \phi_{ij}}{\sin \phi_{ij}}{-\sin \phi_{ij}}{\cos \phi_{ij}}
\cVector {\sigma_{j}^{x} }{ \sigma_{j}^{y}}.
\ee
which may be considered as a twisted $XY$ model.

Our  motivation for this particular quantum metamaterial is a quantum circuit realisation fabricated using and array of superconducting qubits. This justifies our chosen Hamiltonian as it is of the same form as those currently used to model such circuits. It is, however, interesting to note that it would be possible to construct an equivalent system of interacting fermions (this can  be achieved via a Jordan-Wigner or similar transform~\cite{PhysRevA.65.042323}). The results presented here in terms of the states of a qubit can, therefore, also be viewed in terms of the motion of fermions. Such a process might, for example, take the form of electrons propagating in a lattice of quantum dots in the presence of a magnetic flux which is precisely the case of mesoscopic transport mentioned previously.

\begin{figure}[!tb]
\begin{center}
\begin{tikzpicture}[scale=1.0]
\pgftext[bottom,left]{%
\includegraphics[width=0.45\textwidth]{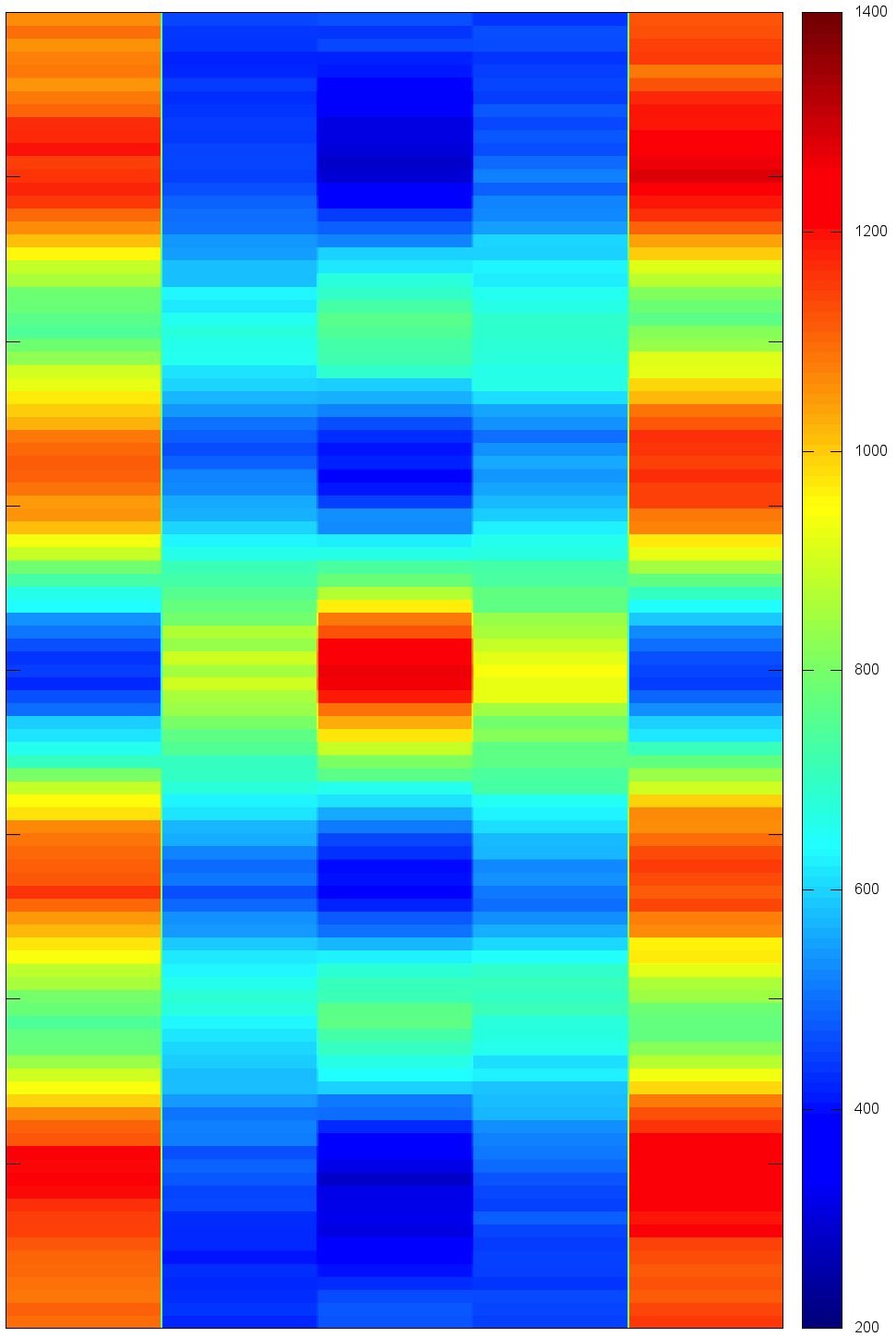}%
}%

\node (mpi) at (-0.25,0.1) {$-\pi$};
\node (mpi2) at (-0.25,3) {$- \dfrac{\pi}{2}$};
\node (pi2) at (-0.25,9) {$ \dfrac{\pi}{2}$};
\node (pi) at (-0.25,6) {$0$};
\node (zero) at (-0.25,12.0) {$\pi$};

\node[circle, shading=ball, ball color=magenta] (a) at (3.5,-2.5) 	{1};
\node[circle, shading=ball, ball color=cyan] (b) at (4.9,-2.0) 			{2};
\node[circle,draw, shading=ball, ball color=green] (c) at (6.3,-1.5) 	{3};

\node[circle, shading=ball, ball color=cyan] (d) at (2.1,-2.0) 			{4};
\node[circle, shading=ball, ball color=cyan] (e) at (3.5,-1.5) 		{5};
\node[circle,draw, shading=ball, ball color=green] (f) at (4.9,-1) 	{6};

\node[circle,draw, shading=ball, ball color=green] (g) at (0.7,-1.5) 	{7};
\node[circle,draw, shading=ball, ball color=green] (h) at (2.1,-1) 	{8};
\node[circle,draw, shading=ball, ball color=green] (i) at (3.5,-0.5) 	{9};

\draw[blue,directed,ultra thick] (a) -- (b) ;
\draw[double, color = black] (b) -- (c) ;

\draw[blue,directed,ultra thick] (d) -- (a) ;
\draw[blue,directed,ultra thick] (b) -- (e) ;
\draw[double, color = black] (c) -- (f) ;

\draw[blue,directed,ultra thick] (e) -- (d) ;
\draw[double, color = black] (e) -- (f) ;

\draw[double, color = black] (d) -- (g) ;
\draw[double, color = black] (e) -- (h) ;
\draw[double, color = black] (f) -- (i) ;

\draw[double, color = black] (g) -- (h) ;
\draw[double, color = black] (h) -- (i) ;
\end{tikzpicture}
\end{center} 
\caption{(color online) We represent along the abscissa each of the qubits to which the output measurement devices are coupled, i.e. qubits; 7, 8, 9, 6 \& 3 (as indicated by the schematic beneath the plot). On the ordinate axis we show the phase $\phi$ for the coupling term $\Xi(\phi)$ of \Eq{eq:coupling}. In the $z$ direction (indicated through the colormap) we show the total number of counts in each detector over ten different solutions to~\Eq{eq:jumps}. Here we have used the following material and detector parameters; $\mu_{ij}=\half$ (for the connected qubits) and $\gamma_\mathrm{in}=\gamma_\mathrm{out}=1$ (we note that results of a similar form are found for $\mu_{ij}=-\half$).}
\label{fig:counts}
\end{figure}
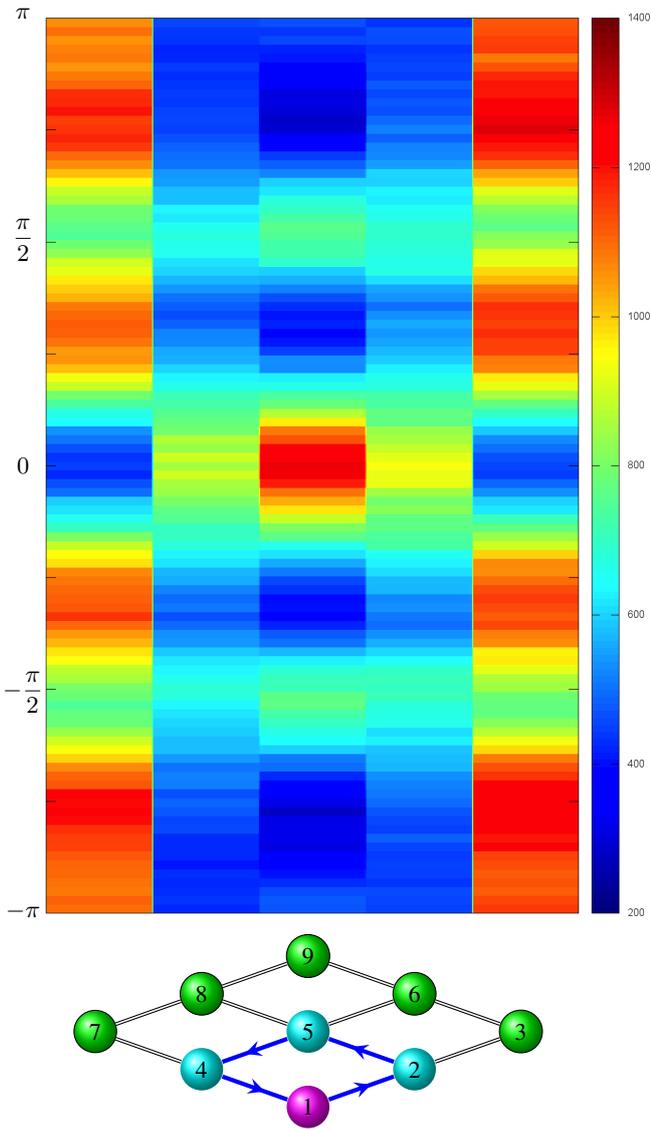

In order to characterise the properties of our proposed quantum metamaterial we need to generate some kind of excitation flow through the material. Hence, we will need an input, which we arbitrarily have coupled to qubit 1. Moreover, we also need to measure the output of this flow on the outer edges of the material, that is, qubits 7, 8, 9, 6 and 3 (reading left to right around the opposite edges to qubit 1). We have chosen to do this through the introduction of quantum jumps based measurement devices (as the outcome of a measurement is clear, either a jump is measured or it is not, see for example~\cite{PhysRevE.72.066209}). In effect, we measure in  and create excitations on node 1 in~\Fig{fig:schematic} and measure out and destroy excitations on the opposing edges (nodes 7, 8, 9, 6 and 3). 

The quantum jumps model is an unravelling of the master equation corresponding to the irreversible emission/absorption of absorbed/emitted excitations over very short time scales. The equation for this unravelling takes the form of stochastic It\^o increment equation for the state vector according to
\bal{eq:jumps}
\ket{\ud \psi}&=&
-\frac{i}{\hbar}H\, \ket{\psi} \ud t 
- \half \sum_{j\in B}\left[L_{j}^{\dag}L_{j}-\EX{L_{j}^{\dag}L_{j}} \right] \ket{\psi}\ud t \nonumber \\
&&+ \half \sum_{j \in B}\left[\frac{L_{j}}{\EX{L_{j}^{\dag}L_{j}}}-1 \right] \ket{\psi}\ud N_{j}
\ea
where  $B=\{1,3,6,7,8,9\}$ and $\ud N_{j}$ is a Poissonian noise process satisfying $\ud N_j\, \ud N_k =\delta_{jk}\ud N_j$, $\ud N_j \ud t =0$ and $\overline{\ud N_j} = \EX{L_{j}^{\dag}L_{j}}\, \ud t$. That is, jumps occur randomly at a rate that is determined by \EX{L_{j}^{\dag}L_{j}}. Here the Lindblad associated with the input node  is $L_\mathrm{1}=\gamma_\mathrm{in}\sigma_{1}^{+}$ and with the output nodes is
$L_\mathrm{i}=\gamma_\mathrm{out}\sigma_{i}^{-}$ for $i\in\{3,6,7,8,9\}$ and zero otherwise.

In order to take account of the statistical nature of unravelling's of the master equation we have, in the results that follow, summed the counts measured over ten trajectories. For each trajectory we integrated the counts measured over a reasonable duration of time ($0\leq t \leq 1000$). These two factors together are sufficient to average out most of the statistical fluctuations that arise when using quantum trajectories methods (we are confident that the slight asymmetries that remain in our results are due to the remaining fluctuations).

On inspection of the results presented in~\Fig{fig:counts} we see several points of interest. 
From the coupling term in ~\Eq{eq:coupling} we first observe that we would expect our results to be symmetric about $\phi=0$ as is indeed the case. If, together with this coupling term, we also examine the loop between nodes $1\rightarrow2\rightarrow5\rightarrow4\rightarrow1$ of~\Fig{fig:schematic}. Here the phases are applied to impose $\phi$ with a hop along the arrow and $-\phi$ in the opposite direction. We see that reversing all the arrows is equivalent to changing the sign of $\phi$ and equivalent to a reflection of the lattice about the axis going through sites 1,5 and 9. Reflecting this symmetry we see that the results of ~\Fig{fig:counts} are invariant on changing $\phi$ to $-\phi$ and simultaneously swapping labels between 3 and 7 and between 6 and 8. 

For certain values of the controlling parameter (phase) $\phi$ we see behaviour that is consistent with that of a quantum birefringent material acting as a beam splitter. We see this from $\phi=\pm\pi$ to around $\pm {3 \over 4}\pi$ (where the effect is strongest) where we see that the vast majority of counts are measured on qubits 7 and 3. This pattern is again repeated around $\phi=\pm {1 \over 4}\pi$. Although not conclusive, the possibility that the material is acting as a quantum birefringent beam splitter  is supported by the calculation of the second order correlation coefficient 
$$
g^{(2)}=1+\frac{\EX{\Delta N_{7}\Delta N_{3}}}{\EX{N_{7}}\EX{N_{3}}}
$$
between the 7th and 3rd qubit at $\phi=3\pi/4$ which we have determined to be approximately $0.92$ which, by analogy with~\cite{PhysRevLett.108.263601}, is indicative of non-classical scattering by the material (here, $N_{i}$ is the record, as a time series, of photons counted at each detector).
In contrast, at $\phi=0$ we see that counts are measured in a bell shape centred around qubit 9. Here it could be argued that this behaviour corresponds to propagation of a signal as a beam across the diagonal of the quantum metamaterial.

Finally we note that at $\phi=\pm {1 \over 2}\pi$ and approximately $\phi=\pm {1 \over 8}\pi$ the distribution of  counts is approximately flat over all the output qubits. At these points the material is somewhat equally opaque across all the detectors with a quantum analogue of electromagnetically induced transparency occurring for the other possible values of $\phi$.

In conclusion, we have presented a model of a two-dimensional quantum metamaterial whose behaviour is tuneable. For different values of the control phase our results are consistent with (i) a quantum birefringent beam splitter (ii) reduced transmission plain wave propagation and (iii) a beam directed along the leading diagonal. This system contains many degrees of freedom and the analysis we have presented here only just begins to scratch the surface of what we believe quantum metamaterials will be capable of. Some example applications, that are beyond the scope of this work, might be demonstrating possible violations of Bell's inequalities for the material in its ``birefringent'' state (e.g. at $\phi=\pm 3 \pi/4$) or further modifying the materials properties by applying the control flux in  a manner that breaks the symmetry we assume in this \PAPER.

\begin{acknowledgments}
MJE, RW, SES and AMZ acknowledge that this publication was made possible through the support of a grant from the John Templeton Foundation; the opinions expressed in this publication are those of the authors and do not necessarily reflect the views of the John Templeton Foundation.

\end{acknowledgments}

\bibliography{ref}

\end{document}